\newcommand{\be}{\begin{equation}}
\newcommand{\ee}{\end{equation}}

\documentclass[prl,twocolumn,showpacs,floatfix]{revtex4}

\usepackage{graphicx}

\begin{document}

\title{Quantum criticality, particle-hole symmetry, and duality \\ of the
plateau-insulator transition in the quantum Hall regime.}

\author{A.M.M. Pruisken}
\email{pruisken@science.uva.nl}
\affiliation{Institute for
Theoretical Physics, University of Amsterdam, Valckenierstraat 65,
1018 XE Amsterdam, The Netherlands}

\author{D.T.N. de Lang}
\author{L.A. Ponomarenko}
\author{A. de Visser}
\affiliation{Van der Waals-Zeeman Institute, University of
Amsterdam, Valckenierstraat 65, 1018 XE Amsterdam, The
Netherlands}

\begin{abstract}
We report new experimental data on the plateau-insulator transition in the quantum
Hall regime, taken from a low mobility InGaAs/InP heterostructure.
By employing the fundamental symmetries of the quantum transport problem
we are able to
disentangle the universal quantum critical aspects of the
magnetoresistance data (critical indices and scaling functions)
and the sample dependent aspects due to macroscopic
inhomogeneities. Our new results and methodology indicate
that the previously established experimental value for the critical
index ($\kappa = 0.42$) resulted from an admixture of both
universal and sample dependent behavior. A novel, non-Fermi liquid value
is found ($\kappa = 0.57$) along with the leading corrections to scaling.
The statement of self-duality under the Chern Simons flux attachment
transformation is verified.
\end{abstract}


\maketitle

More than ten years ago, H.P.Wei \textit{et al.}
\cite{experiments} demonstrated experimentally that the plateau
transitions in the quantum Hall regime display all the features of
a conventional, disorder induced metal-insulator
transition~\cite{freepart}. The transport data, taken from low
mobility InGaAs/InP heterostructures at low temperatures $T$, were
shown to depend on the single scaling variable
$T^{-\kappa}\Delta\nu$ only. Here, $\Delta\nu$ represents the
filling fraction of the Landau band $\nu$ relative to the critical
value $\nu_c \approx$ half-integer. The quantity of central
physical interest is the exponent $\kappa$. This represents the
ratio of two independent critical indices $\kappa =
p/2\tilde{\nu}$. Here, the exponent $p$, originally identified as
the inelastic scattering time exponent, determines the relation
between the phase breaking length $l_\phi$ and temperature $T$
according to $l_\phi \propto T^{-p/2}$. The $\tilde{\nu}$ is the
exponent of the correlation length $\xi$ that diverges at the
Landau band center according to $\xi =
|\Delta\nu|^{-\tilde{\nu}}$. The quoted best experimental value
for $\kappa$ is given by $\kappa \approx
0.42$~\cite{experiments,hwang}.

However, even currently the precise meaning of the exponent
$\kappa$ and the very nature of the quantum critical phase remain
a major topic of both experimental and theoretical research. From
the theoretical side, a microscopic theory for localization and
Coulomb interaction effects has been developed only
recently~\cite{paperIII,twoloop,pbb}. The theory now shows that the physical
observable, associated with the exponent $p$, is none other than
the specific heat of the electron gas $c_v \propto T^p$. From the
experimental point of view, there are still major complications in
approaching the true $T=0$ asymptotics of scaling. One of the main
difficulties in probing quantum criticality in the quantum Hall
regime is that the experiment must be performed on samples where
the scattering is mainly provided by short-ranged potential
fluctuations. Slowly varying potential fluctuations, like those
present in the GaAs/AlGaAs heterostructure, give rise to
cross-over effects that dramatically complicate the observability
of quantum criticality in the quantum Hall regime~\cite{paperIII}.

The effect of macroscopic sample inhomogeneities, which are
inherent to the experiments, on the critical behavior of the
plateau transitions is still not known. For example, there are
problems in experimentally establishing a universal value for the
critical resistivities or conductivities at the plateau-plateau
(PP) transitions. A deeper understanding of this issue is
essential for the theory of quantum transport where the effects of
mesoscopic conductance fluctuations on the macroscopic transport
parameters have remained a subject of debate~\cite{twoloop}.

The problem of macroscopic sample inhomogeneities was recently
addressed by van Schaijk \textit{et al.}~\cite{schaijk} who
investigated the plateau-insulator (PI) transition in the lowest
Landau level. The experiments were conducted on the same
InGaAs/InP sample which was used previously by Hwang \textit{et
al.} in their study of the plateau-plateau (PP)
transitions~\cite{hwang}.

In this Letter we report the existence of universal scaling
functions for the transport parameters, as well as a corrected
value for the universal exponent $\kappa$, obtained from new
measurements and by reinvestigating the experiments by van
Schaijk \textit{et al.}~\cite{schaijk} on the PI transition. The
very special features of the lowest Landau level data permit the
observation of the universal aspects of quantum criticality that
have so far not been accessible.

It is important to emphasize that in transport experiments in the
quantum Hall regime, one is generally faced with the problem of
how to extract a local resistivity tensor $\rho_{ij}$ from the
measured macroscopic resistances $R_{ij}$. The experiment is
usually conducted on samples which are prepared in a Hall bar
geometry. The experimentally measured quantities are in this case
$R_{xx}$ and $R_{xy}$ (we take the x-direction as the direction of
the electrical current). Assuming that the electron gas is
homogeneous then the resistivity components $\rho_{xx}$ and
$\rho_{xy}$ are simply obtained as
\be
R_{xx} = \frac{L}{W} \rho_{xx} \;\;\;\; R_{xy} = \rho_{xy}.
\ee
Here, $L/W$ is a geometrical factor with $L$ denoting the
``length" and $W$ the ``width" of the Hall bar respectively.

It is well known, however, that experiments usually do not provide
unique and well-defined values of $\rho_{xx}$ and $\rho_{xy}$.
Slightly different results are obtained by measuring on different
pairs of contacts on the Hall bar, for example, or by reversing
the direction of the magnetic field $B$. This means that the
homogeneity assumption is violated for realistic samples and the
relation between the resistivity tensor and the macroscopic
observables is more complex and non-local in general.

For a general understanding of the quantum critical phenomenon it
is important to know how macroscopic inhomogeneities are different
in effect from, say, anisotropies in the transport parameters that
originate from the microscopics of the sample. In both cases one
would generally require an experimental knowledge of \textit{all
four} components of the resistance tensor $R_{ij}$ rather than
only the quantities $R_{xx}$ and $R_{xy}$ as obtained from the
experiments on the Hall bar geometry.

To discuss the case of an anisotropic resistivity tensor we
consider a rectangular sample of size $L \times W$. The $R_{ij}$
and the $\rho_{ij}$ are related in the standard manner and we
split the latter in pieces that are \textit{symmetric} and
\textit{antisymmetric}
\be
\rho_{ij} (B) = \rho_{ij}^s (B) +\epsilon_{ij} \rho_H (B),
\label{eq:rho_separate}
\ee
\noindent{where} $\rho_{ij}^s (B) = \rho_{ij}^s (-B)$ and $\rho_H
(B) = -\rho_H (-B)$. We may take the $\rho_{ij}^s$ to be, at the
same time, a \textit{symmetric tensor} whereas the quantity
$\epsilon_{ij}$, multiplying the Hall resistance $\rho_H$, is
\textit{antisymmetric}. Next, we decompose the $\rho_{ij}^s$ in an
\textit{isotropic resistivity} $\rho_0$ and an anisotropic part
$S_{ij}$ which we name the \textit{stretch tensor}
\be
\rho_{ij}^s = S_{ij} \rho_0 , \;\;\; \rho_0 = \sqrt{\rho_{xx}^s
\rho_{yy}^s -(\rho_{xy}^s )^2} .\label{eq:rho_sym}
\ee

The stretch tensor $S$ can be represented in terms of an angle of
rotation ($\phi$) and a scale factor ($\lambda$)
\begin{eqnarray}
S_{ij} = \left[\matrix{cos\phi & -sin\phi \cr sin\phi &
cos\phi}\right] \left[\matrix{\lambda & 0 \cr 0 &
\lambda^{-1}}\right]\left[\matrix{cos\phi & sin\phi \cr -sin\phi &
cos\phi }\right]\label{eq:S_ij}
\end{eqnarray}

In experiments on quantum transport one is usually interested in
the temperature ($T$), frequency ($\omega$) and magnetic field
($B$) dependence of the $\rho_0$ and $\rho_H$ alone. In
particular, quantum criticality in the quantum Hall regime implies
that the local resistivity components, at sufficiently low $T$,
become functions of a single scaling variable only
\be
\rho_0 (B,T)  =  \rho_0 ( T^{-\kappa} \Delta\nu) , \;\; \rho_H
(B,T)  =  \rho_H ( T^{-\kappa} \Delta\nu) .
\ee

The functions $\rho_0 (X)$ and $\rho_H (X)$ are given as regular
(differentiable) functions of (small) $X$, with
$X=T^{-\kappa}\Delta\nu$. The concept of quantum criticality is
completely independent of the anisotropy components $S_{ij}$
which, in principle, can be arbitrary ($S_{ij}$, in the context of
the renormalization group, is \textit{irrelevant}). However, if
$S_{ij}$ turns out to play, for some reason, a significant role in
the $T$ and $B$ dependence of the raw experimental data, then it
is necessary to disentangle $S_{ij}$ and $\rho_0$, $\rho_H$ before
the quantum critical behavior of the electron gas can be studied.
In brief we may write
\be
\rho_{ij} (B,T) = S_{ij} (B,T) \rho_0 (X) + \epsilon_{ij} \rho_H
(X) . \label{eq:rho}
\ee

The stretch tensor, unlike the scaling functions $\rho_0 (X)$ and
$\rho_H (X)$, may have a different physical significance and a
distinctly different symmetry, depending on the specific problem
that one is interested in. For the strongly disordered, low
mobility heterostructures which are of interest to us, the
quantity $S_{ij}$ can only have a geometrical significance due to
imperfections and inhomogeneities of the Hall bar. In what follows
we apply the result, Eq.~\ref{eq:rho}, to the special case of the
PI transition and show that the very notion of a stretch tensor
dramatically alters the interpretation of the experiments on
quantum criticality in the quantum Hall regime.

$\bullet$ \textit{Particle-hole symmetry.}
The solution lies, for the major part, in recognizing that quantum
criticality in the quantum Hall regime exhibits a fundamental
symmetry~\cite{freepart,twoloop} which we call particle-hole
symmetry. To show what it means we invert the resistivity tensor,
Eq.~\ref{eq:rho_sym}, to obtain the conductivity tensor. This
tensor can be written as
\be
\sigma_{ij} (B,T) = {\tilde S}_{ij} (B,T) \sigma_0 (X) -
\epsilon_{ij} \sigma_H (X). \label{eq:sigma_ij}
\ee

Here, ${\tilde S}_{ij}$ is the same as in Eq.~\ref{eq:S_ij} except
that $\lambda$ is now replaced by $1/\lambda$ and $\phi$ by
$-\phi$. Furthermore
\be
\sigma_0 = \frac{\rho_0 (X)}{\rho_0^2 (X) +\rho_H^2 (X)} ,~
\sigma_H = \frac{\rho_H (X)}{\rho_0^2 (X) +\rho_H^2 (X)} .
\label{eq:sigma}
\ee

The statement of particle-hole symmetry now implies that at
sufficiently low $T$, the quantities $\sigma_0$ and $\sigma_H$ of
the $PI$ transition satisfy the relation
\be
\sigma_0 (X) =\sigma_0 (-X) , \;\;\;\; \sigma_H (X) =1- \sigma_H
(-X) . \label{eq:parthole}
\ee

It is specifically this relation (Eq.~\ref{eq:parthole}) that
enables us to unravel the problem.

$\bullet$ \textit{The experiment.}
The remainder of this Letter reports a detailed study of the
resistivity tensor taken from the lowest Landau level of an
InP/InGaAs heterostructure~\cite{newexpt}. Our sample and
experimental set-up are identical to those of van Schaijk
\textit{et al.}~\cite{schaijk}.


\begin{figure}[htb]
\includegraphics[height=7cm]{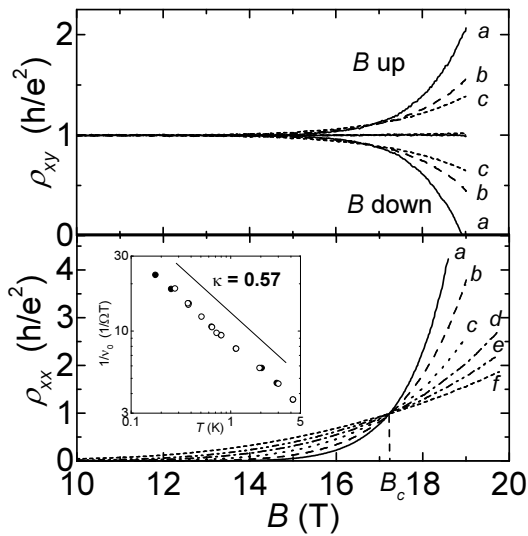}
\caption{\label{fig:1} Data for $\rho_{xx}$ and $\rho_{xy}$ with
varying $B$ and for opposite field directions, taken from an
InGaAs/InP heterojunction ($n = 2.2\times 10^{11}$cm$^{-2}, \mu =
16000$ cm$^2$/Vs). The $a,b,...f$ indicate $T= 0.37, 0.62, 1.2,
1.9, 2.9$ and $4.2 K$ respectively. Inset: $1/\nu_0$ versus
$T$ for the PI transition. The closed and open symbols refer to
opposite directions of the $B$ field. }
\end{figure}

Fig.~\ref{fig:1} shows the results for sweeps in both directions
of the $B$ field for different values of $T$. Upon reversing the
direction of $B$ at constant $T$ we find that the $\rho_{xx}$ data
for the $PI$ transition remain unchanged. The $\rho_{xy}$ data,
however, are strongly affected and the results for opposite $B$
fields display a symmetry about the plateau value $|\rho_{xy} |=
h/e^2$ for low values of $T$ ($T < 1.2K$).

This important phenomenon has been observed before, on samples
that do not provide access to quantum
criticality~\cite{fieldflip}. However, in this Letter we probe
fundamental aspects of universality that cannot be seen otherwise.

Following Fig.~\ref{fig:1} we split the $\rho_{xy}$ data, at
constant $T$, in a \textit{symmetric} and an
\textit{antisymmetric} piece (from now onwards we work in units of
$h/e^2$ )
\be
\rho_{xy} (B) = \rho_H (B) + \rho_{xy}^s (B) , \label{eq:rho_xy}
\ee

\noindent{where} $\rho_H (B) = -\rho_H (-B)$ and $\rho_{xy}^s (B)
= \rho_{xy}^s (-B)$. Since we also have $\rho_{xx} (B)
=\rho_{xx}^s (B) = \rho_{xx}^s (-B)$ we conclude that the
resistivity tensor as a whole can be split, like
Eq.~\ref{eq:rho_separate}, in a symmetric piece $\rho_{ij}^s (B) =
\rho_{ij}^s (-B)$ and an anti-symmetric piece $\rho_H (B) =
-\rho_H (-B)$. This is a very special feature of the $PI$
transition which is generally violated by the $PP$ transitions.

Notice that the Hall resistance $\rho_H$, at low enough $T$,
remains quantized throughout the $PI$ transition. Following
Eq.~\ref{eq:parthole} we obtain an important statement made on the
scaling function $\rho_0 (X)$
\be
\rho_0 (X) = \rho_0^{-1} (-X).\label{eq:duality}
\ee

To establish the significance of this result, we next discuss the
experimental data in more detail.

$\bullet$ \textit{The $\rho_{xx}$ data.}
These data display a fixed point at $B_c = 17.2$ T ($\nu_c=0.55$)
where the various isotherms intersect (Fig.~\ref{fig:1}). A
detailed analysis of the $B$ and $T$ dependence of the $\rho_{xx}$
data indicates that this fixed point represents a true critical
fixed point. The following expression has been extracted
\be
\rho_{xx} (B,T) = \rho_c e^{-X-O(X^3 )} , \;\; X= \Delta\nu /\nu_0
(T).\label{eq:rho_xx}
\ee
Here, $\Delta\nu = n_0 ({n}^{-1} - {n_c}^{-1} )$ with $n_0$
denoting the electron density, $n=\frac{eB}{hc}$ equals the
density of the filled Landau band and $n_c =\frac{eB_c}{hc}$. The
quantity $\nu_0 (T)$ stands for the algebraic expression $\nu_0
(T) = (T/T_0 )^{\kappa}$ (with $T_0 = 52K$, see inset
Fig.~\ref{fig:1}). The critical exponent $\kappa = 0.57 \pm 0.03$
has the same numerical value, within the experimental error, as
the one found by van Schaijk \textit{et al.}~\cite{schaijk}.
The numerical value of
the amplitute $\rho_c$ has been determined to be $\rho_c=
1.0 \pm 0.1$. The large error in $\rho_c$ is mainly
due to the experimental difficulties in measuring the
geometrical factor $L/W$ of the Hall bar.
Next, the term $O(X^3)$ in the exponential indicates that
the next-to-leading term is of order $X^3$. It is difficult,
however, to obtain a reliable estimate for the coefficient in the
limit where $T \rightarrow 0$.

Eqs~\ref{eq:rho_xy},
\ref{eq:duality} and \ref{eq:rho_xx} permit us to extract the
following scaling results for the $PI$ transition in this limit

\be
\rho_0 (X) = {\rho_{xx} (B,T)}/{\rho_c }= e^{-X-O(X^3
)}~;~~~\rho_H = 1. \label{eq:rho0_rhoH}
\ee

In the remainder of this Letter we extend these results to include
both the stretch tensor $S_{ij}$ and the corrections to scaling.

$\bullet$ \textit{The $\rho_{xy}^s$ data.}
The different isotherms $\rho_{xy}^s$ display, just like
$\rho_{xx}$, a fixed point. However, the critical value $B_c$ is
slightly different for the $\rho_{xy}^s$ and the $\rho_{xx}$ data.
Since the measurements were conducted on different parts of the
sample, the difference ($\delta B_c$) in $B_c$ can be attributed
to small gradients in the electron density. The experimental data
can be represented as follows
\be
\rho_{xy}^s (B) =\epsilon ~ \rho_{xx} (B-\delta B_c) = \epsilon
~e^{\frac{\delta \nu_c}{\nu_0 (T)}} ~\rho_{xx} (B) .
\label{eq:rhoXY_symm}
\ee
Here, $\delta \nu_c$ denotes the difference in critical filling
factors. The coefficient $\epsilon$ is found to be small
($\epsilon = 0.1$) and independent of $B$ and $T$.

$\bullet$ \textit{The stretch tensor.}
In order to demonstrate that Eq.~\ref{eq:rhoXY_symm} is the
combined result of a misalignment of sample contact and density
gradients, we have computed $S_{ij}$ for an imperfect system where
the Hall bar geometry is that of a parallelogram obtained by
rotating the $Y$-axis of the $L \times W$ rectangular system over
a small angle $\theta$. To represent the small gradients in the
electron density we have assumed that the resistivity components
are given as in Eq.~\ref{eq:rho0_rhoH} but with a spatially
dependent critical filling fraction $\nu_c \rightarrow \nu_c (x,y)
= \nu_c + ax + by$. The results for $\rho_{ij}$ are precisely of
the form of Eq.~\ref{eq:rho} with the stretch tensor now given
by~\cite{newexpt}
\begin{eqnarray}
S_{ij} (B,T) = \left[\matrix{\frac{1}{cos\theta} &
-tg\theta~e^{\frac{\delta\nu_y}{\nu_0(T)}} \cr
tg\theta~e^{\frac{\delta\nu_x}{ \nu_0(T)}} & \frac{1}{cos\theta}
}\right] . \label{eq:stretchtensor}
\end{eqnarray}

Here, $\delta\nu_x =\pm aL/2$ and $\delta\nu_y =\pm bW/2$ are the
uncertainties in the critical filling fractions $\nu_c$ in the $x$
and $y$ directions respectively. Eq.~\ref{eq:stretchtensor}
contains all the sample dependence in the transport parameters due
to macroscopic inhomogeneities and compares well with the
experimental results. On the basis of Eqs~\ref{eq:rho_xx} and
~\ref{eq:rhoXY_symm} we identify $\delta\nu_x = \delta\nu_c$,
$sin\theta = \epsilon$ and ${cos\theta}= \rho_c^{-1}$. The
quantity $\delta\nu_c /\nu_0 (T)$ is less than 0.2 in the regime
of experimental $T$. This is consistent with the condition $\left[
\delta\nu_{x,y} / \nu_0 (T) \right]^2 \ll 1$ under which
Eq.~\ref{eq:stretchtensor} is valid. Furthermore, the numerical
values $\theta \approx \epsilon =0.1$ and $\rho_c = 1.0$ are quite
reasonable and well within the range of experimental
uncertainties.

\begin{figure}[htb]
\includegraphics[height=7cm]{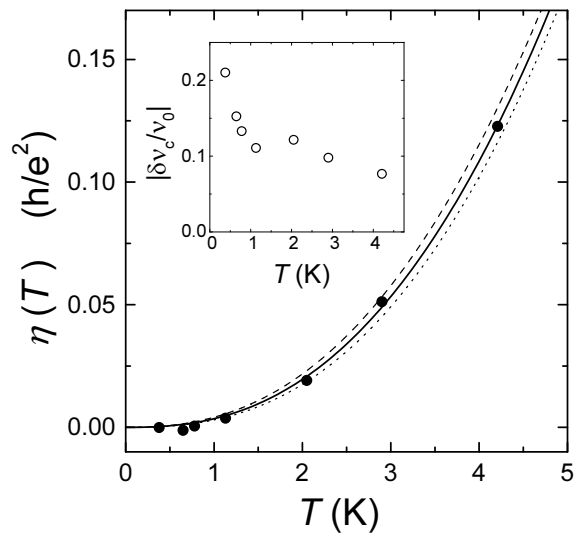}
\caption{\label{fig:2} The powerlaw correction $\eta(T)$ to the
exact quantization of $\rho_H$, Eq.~\ref{eq:eta}. Closed circles:
data for $\eta (T)$ with $\nu=\nu_c$ and varying $T$. Solid curve:
algebraic fit with $y_\sigma =2.5$ and $T_1 =9.8$K. Away from the
fixed point ($\nu \neq \nu_c$), the extracted value of $y_\sigma$
varies by less than 3\% with varying values of $B$. Dotted curve:
$B=15.8$T and varying $T$; dashed curve: $B=19.0$T and varying
$T$. Inset: $|\delta\nu_c / \nu_0|$ vs $T$.}
\end{figure}

$\bullet$ \textit{Corrections to scaling.}
To complete the analysis of the resistivity tensor, the small
deviations in $\rho_H$ from exact quantization
are now addressed. The following expression (see Fig.~\ref{fig:2} )
\begin{equation}
\rho_H = 1 +\eta(T) e^{-X} , ~~
\eta (T) = \left( {T / T_1} \right)^{y_\sigma} \label{eq:eta}
\end{equation}
with $y_\sigma = 2.5$ and $T_1 = 9.8K$ provides an accurate description
of the transport data on the quantum critical phase with $T<4K$.
This leads to the most important result of this paper,
Eq.~\ref{eq:sigma}, which can now be extended to include the
corrections to scaling. Working to linear order in $\eta (T)$ we
obtain
\be
\sigma_0  =  \frac{\rho_0 }{\rho_0^2 + 1 + 2\eta  \rho_0 }
,~\sigma_H  =  \frac{1+\eta \rho_0 }{\rho_0^2 + 1 + 2\eta \rho_0}.
\label{eq:correction}
\ee
These final results, when plotted as $T$-driven flowlines
in the $\sigma_0$, $\sigma_H$
conductivity plane, display all the fundamental features of
scaling as predicted by the renormalization theory
\cite{freepart,twoloop}.

In summary, this is the first time that both universal scaling
functions and the exponent $\kappa$ have been simultaneously
extracted for the $PI$ transition. We emphasize that these results
verify quantum criticality, particle-hole symmetry as well as the self-duality
of the transport
parameters in accordance with the Chern Simons gauge theory for
abelian quantum Hall states~\cite{paperIII,kivel}.
%
The new experimental value that we have obtained for the exponent
($\kappa = 0.57 \pm 0.03$) is in conflict with the Fermi liquid
ideas that have been frequently proposed over the
years~\cite{believe}. Since $p$ is bounded by
$1<p<2$~\cite{twoloop}, the experimental value $\kappa = 0.57 \pm
0.03 $ implies that the correlation length exponent is bounded by
$0.9<\nu<1.8$. The Fermi liquid value equals $\nu =2.3$ as
obtained from numerical work~\cite{kramer}. We may therefore
conclude that the Coulomb interaction problem and the free
electron gas lie in different universality
classes~\cite{twoloop,pbb}.

The research was supported in part by FOM and INTAS (Grant
99-1070).

\end{document}